\documentclass[reprint,superscriptaddress,amsmath,amssymb,aps,nofootinbib]{revtex4-1}

\usepackage{graphicx}
\usepackage{dcolumn}
\usepackage[french,english]{babel}
\usepackage[utf8]{inputenc}
\usepackage{enumitem} 
\usepackage{revsymb}
\usepackage{xcolor}
\usepackage{bbold}
\usepackage{bm}
\usepackage[colorlinks]{hyperref}

\hypersetup{
	colorlinks   = true,
	citecolor    = gray,
	linkcolor	 = blue,
	urlcolor = black
}

\usepackage{newfloat}
\usepackage{etoolbox}
\AtEndEnvironment{algorithm}{\noindent\hrulefill\par\nobreak\vskip-8pt}

\usepackage{array}
\newcolumntype{L}[1]{>{\raggedright\let\newline\\\arraybackslash\hspace{0pt}}m{#1}}
\newcolumntype{C}[1]{>{\centering\let\newline\\\arraybackslash\hspace{0pt}}m{#1}}
\newcolumntype{R}[1]{>{\raggedleft\let\newline\\\arraybackslash\hspace{0pt}}m{#1}}

\newcommand{\main}{\noindent\hspace*{10pt}\ignorespaces}

\makeatletter
\renewcommand{\fnum@figure}{FIG. \thefigure}
\makeatother

\begin{document}
	
	\title{A driven-dissipative quantum Monte Carlo method for open quantum systems}
	
	\author{Alexandra Nagy}
	\affiliation{Institute of Physics, Ecole Polytechnique Fédéral de Lausanne (EPFL), CH-1015, Lausanne, Switzerland}
	\author{Vincenzo Savona}
	\affiliation{Institute of Physics, Ecole Polytechnique Fédéral de Lausanne (EPFL), CH-1015, Lausanne, Switzerland}

	\begin{abstract}
		We develop a real-time Full Configuration Interaction Quantum Monte Carlo approach for the modeling of driven-dissipative open quantum systems. The method enables stochastic sampling of the Liouville-von-Neumann time evolution of the density matrix, thanks to a massively parallel algorithm, thus providing estimates of observables on the non-equilibrium steady state. We present the underlying theory, and introduce initiator technique and importance sampling to reduce the statistical error. Finally, we demonstrate the efficiency of our approach by applying it to the driven-dissipative two-dimensional XYZ spin model on lattice.
	\end{abstract}
	
	\maketitle
	
	\section{\label{sec:intro} Introduction}
	
	\main The study of the nonequilibrium dynamics of many-body open quantum systems has attracted increasing attention in recent years, due to the progress in several experimental areas, including ultracold atomic gases, trapped ions, and superconducting circuits \cite{carusotto_quantum_2013,hartmann_quantum_2016-1,le_hur_many-body_2016,noh_quantum_2017}. A common feature of these systems is the coupling to an external environment in the form of coherent or incoherent input and output channels. The time evolution of the system is then governed by the Liouville-von-Neumann equation which -- in the case of stationary external conditions -- typically drives it into a nonequilibrium steady state (NESS). Here, the competition between the coherent and incoherent dynamics gives rise to a multitude of novel phenomena, including nonequilibrium dissipative phase transitions \cite{kessler_dissipative_2012,diehl_quantum_2008}.\\
	\main Generally, the steady state density matrix can be obtained in two different ways. First, one can integrate the time evolution until the stationary state is reached, or second, one can find the solution associated with the null eigenvalue of the Liouvillian super-operator. The theoretical description of open many body systems represents a major challenge, and in spite of the numerous improvements, the numerical modeling can be handled only for small system sizes. While several studies have been restricted to mean-field approximations \cite{nissen_nonequilibrium_2012,jin_photon_2013,tomadin_signatures_2010}, in the case of one-dimensional systems highly accurate results were obtained via the density matrix renormalization group (DMRG) technique \cite{white_density_1992,schollwock_density-matrix_2005,de_chiara_density_2008}, and the equivalent variational approach based on the matrix product state (MPS) ansatz \cite{verstraete_matrix_2008,schollwoeck_density-matrix_2011}. Alternatively, a stochastic method, i.e the Monte Carlo Wave Function (MCWF) \cite{gardiner_wave-function_1992,dum_monte_1992}, can be used which unravels the system-bath interaction onto a stochastic process that adds to the unitary Hamiltonian dynamics and to the effective damping terms. Furthermore, a spatial renormalization approach -- the corner space renormalization method -- was recently proposed, which relies on an ad hoc spatial decimation protocol for the density matrix. \cite{finazzi_corner_2015,rota_critical_2017}.\\
	\main For closed, Hamiltonian systems, various quantum Monte Carlo approaches have been the election tools to stochastically sample system properties, both at zero and finite temperature \cite{kolorenc_applications_2011,foulkes_quantum_2001}. A class of methods generally known as projector Monte Carlo (PMC) \cite{umrigar_observations_2015} -- such as diffusion Monte Carlo (DMC) \cite{kosztin_introduction_1996,umrigar_diffusion_1993} and Green's function Monte Carlo (GFMC) \cite{kalos_monte_1962} -- enables modeling the ground state properties at zero temperature by stochastically sampling the time evolution of the imaginary-time Schrödinger equation. However, PMC methods may suffer from the sign problem which results in a computational cost growing exponentially with the system size. Recently, a novel approach called Full Configuration Interaction Monte Carlo (FCIQMC) has been introduced for quantum chemistry simulations \cite{booth_fermion_2009, spencer_sign_2012,booth_approaching_2010,cleland_communications:_2010} and for the simulation of other strongly correlated systems \cite{booth_towards_2013,umrigar_alleviation_2007}. As a projector technique for zero temperature, it has features in common with DMC and GFMC, although a radically different sampling protocol was introduced which has proven to significantly alleviate the sign problem \cite{shepherd_sign_2014,umrigar_alleviation_2007,spencer_sign_2012}. Nevertheless, the sign problem is NP-hard problem \cite{troyer_computational_2005},and is not completely solved by FCIQMC. \\ 
	\main A mutual feature of the Liouvillian dynamics and the imaginary-time Schrödinger equation is the fact, that in the long-time limit, the eigenstate with the smallest-real-part-eigenvalue will dominate. In the Liouvillian case, where the eigenvalues of the Liouvillian super-operator are complex valued with negative real part \cite{albert_symmetries_2014-1}, this corresponds to the null eigenvalue solution -- to the NESS [Table~\ref{tab:tab1}]. It would therefore be natural to apply projector Monte Carlo techniques to the simulation of the NESS. In this case however, we expect the sign problem to be highly relevant, as the complex valued density matrix cannot be expected to possess elements of the same sign only.\\
	\begin{table*}[t!]
		\begin{tabular}{ | C{2.5cm} || C{4.7cm} | C{4.7cm} | }
			\hline
			& Closed system & Open system \\ \hline \hline
			System \newline operator & Hamiltonian \newline $H = H^\dagger$ & Liouvillian \newline  $\mathcal{L}$\\ \hline
			
			Dynamics & Imaginary-time \newline \vspace*{-0.5cm} 
			\begin{equation*} \label{eq:imSch}
				\dot{\psi}(\tau) = -(\hat{H}-E_0)\psi(\tau)
			\end{equation*}\vspace*{-0.5cm}   & Real-time \newline  \vspace*{-0.5cm} 
			\begin{equation*}
				\dot{\hat{\rho}}(t) = \mathcal{L}\hat{\rho}
			\end{equation*}\vspace*{-0.5cm} \\ \hline
			
			Long-time \newline limit & Ground state \newline \vspace*{-0.5cm}
			\begin{equation*} \label{eq:gensol}
				e^{-\tau(\hat{H}-E_0)}: \psi_{in} \xrightarrow[]{\tau \rightarrow \infty} \psi_{0}
			\end{equation*}\vspace*{-0.5cm} &
			Nonequilibrium steady state  \newline \vspace*{-0.5cm} 
			\begin{equation*}
				e^{\mathcal{L}t}: \rho_{in} \xrightarrow[]{t \rightarrow \infty} \rho_{ss} 
			\end{equation*}\vspace*{-0.5cm}  \\ \hline \hline
			
		\end{tabular}
		\caption{\label{tab:tab1} A parallel is drawn between the imaginary-time evolution of closed hamiltonian systems and the real-time evolution of the quantum master equation for open quantum systems. In the case of open systems we assume, here and throughout this work, that a unique steady state exists.
		}
	\end{table*}
	\main In this paper, we develop a real-time FCIQMC approach to open quantum systems, which we call driven-dissipative quantum Monte Carlo (DDQMC). DDQMC shares many of the features of FCIQMC, but it samples the elements of the complex-valued density matrix instead of the wavefunction. The method does not truncate the Hilbert-space and contrary to tensor network methods, its applicability is not bound to the dimensionality of the system. In order to demonstrate the use of DDQMC, we simulate a two-dimensional spin lattice governed by the Heisenberg XYZ Hamiltonian interacting with a dissipative environment. The study of this model has recently attracted interest as an example of a dissipative phase transition resulting from the non-trivial competition between the coherent and incoherent dynamics \cite{rota_critical_2017,rota_dynamical_2017}. The single-site Gutzwiller mean-field study of the system predicts a phase transition from a paramagnetic phase to a magnetically ordered one \cite{lee_unconventional_2013,casteels_gutzwiller_2017}, while a recent analysis showed that this transition should survive in 2D and disappear in case of one-dimensional lattices \cite{jin_cluster_2016}. 
	\\
	\main The paper is organized as follows. In Section~\ref{sec:fciqmc} we give an overview of the original FCIQMC algorithm, setting the basis to the formulation of DDQMC in Section~\ref{sec:ddqmc}. In Section~\ref{sec:initiator} the initiator approach and the importance sampling are introduced. The method is then applied to the XYZ Heisenberg lattice in Section~\ref{sec:results} and the results are compared to those obtained by an optimized exact diagonalization method and by quantum trajectories. We finally discuss the effectiveness of the approach and offer some concluding remarks.
	


	\section{\label{sec:fciqmc} Overview of the FCIQMC algorithm}
	\main We begin by giving a short overview of the FCIQMC method. For a more complete derivation readers are referred to Refs.~\cite{booth_fermion_2009,spencer_sign_2012,shepherd_sign_2014,booth_linear-scaling_2014,blunt_density-matrix_2014}.
	\main In general, PMC methods are stochastic implementations of the power method which aims at computing the expectation values of operators on the dominant eigenstate of the projector. They prove to be particularly useful when the Hilbert-space is so large that the storage of matrix and vector representations becomes computationally unfeasible. PMC techniques get around this memory limitation by storing at any instant in time only a random sample of vector and matrix elements. The expectation values are then computed as time averages. In common with PMC, FCIQMC also performs the long-time integration of the imaginary-time Schrödinger equation. However, unlike PMC, this is achieved with a completely different sampling strategy.\\
	\main Consider the imaginary-time Schrödinger equation (we assume here and in what follows $\hbar = 1$)
	\begin{equation}
	\dot{|\psi\rangle} = -\hat{H}|\psi\rangle\,,
	\end{equation}
	the general solution of the equation is
	\begin{equation}
	|\psi(\tau)\rangle = e^{-\hat{H}\tau}|\psi(\tau=0)\rangle\,.
	\end{equation}
	Once expanded onto the basis spanned by the eigenvectors of the Hamiltonian $\{|\phi_i\rangle\}$, the wavefunction results in a sum of exponentially decaying terms. In order to prevent the ground state component from decaying in the infinite time limit, a constant energy shift $E_0$ can be introduced, where $E_0$ is the ground state energy. Since the value of $E_0$ is unknown in advance, one solves the equation with an arbitrary shift S
	\begin{equation}\label{eq:imSchr}
	\dot{|\psi\rangle} = -(\hat{H}-S\mathbb{1})|\psi\rangle\,.
	\end{equation}
	During the simulated time evolution the value of the $S$ is slowly adjusted in order to maintain a constant normalization and -- at convergence -- provides an estimate of the actual ground state energy $E_0$. \\
	\main FCIQMC stochastically samples the first order Euler approximation of eq.~\eqref{eq:imSchr}. Furthermore, the algorithm works on a discrete basis set $\{|\phi_i\rangle\}$, and the Hamiltonian and the wavefunction are projected onto the space spanned by the basis elements
	
	\begin{equation}
	\psi(\tau) = \sum_{i}^{}c_i^\tau |\phi_i\rangle\,.
	\end{equation}
	
	\noindent The evolution of the expansion coefficients is then governed by
	
	\begin{equation}\label{dyn_eq}
	c_i^{(\tau + \Delta\tau)} = [1 - \Delta\tau(H_{ii}-S)]c_i^{\tau} - \Delta\tau\sum\limits_{j \neq i}^{} H_{ij}c_j^{\tau}\,,
	\end{equation}
	\noindent where $H_{ij}=\langle\phi_i|\hat{H}|\phi_j\rangle$.\\
	\main In order to stochastically represent eq.~\eqref{dyn_eq}, we introduce a fundamental unit called \textit{walker}. Each walker has a sign ($q = \pm1$), and contributes to sampling the amplitude of one of the $|\phi_i\rangle$ basis states. Let $n_i^+$ be the number of walkers with positive sign on a given state and $n_i^-$ that of walkers with negative sign. Then the amplitude of a basis state in the expansion is proportional to the net walker number residing there: $c_i \propto n_i^+-n_i^-$. Starting from an initial distribution, the walkers evolve following a set of rules designed to sample the time evolution of eq.~\eqref{dyn_eq} over a time step. This dynamics is then iterated until convergence is reached. \\
	\main We point out here, that this approach sets an upper bound to the time step. Since the initial state is driven into the dominant eigenstate of the projector, the solution will converge to the ground state only if $|1-\Delta\tau(E_i-S)| \leq 1$ for all eigenvalues $E_i$. This corresponds to the requirement $\Delta\tau \leq 1/\Delta$, where $\Delta$ is the full spectral width of the Hamiltonian under consideration.\\
	\main The rules for evolving the walker population can be summarized as follows. At each time step we loop over the entire walker population and perform the following operations:
	
	\begin{enumerate}[label=(\roman*)]
		
		\item \textit{Spawning:} For a walker residing on site $i$ a connected site $j$ is chosen randomly and a spawning event is made possible with a probability $p(j|i)\propto|H_{ji}|\Delta\tau$ (connected sites are linked by non-zero off-diagonal Hamiltonian elements $H_{ij}$). If the attempt is successful, walkers are born at site $j$ with sign $q_j = \mathrm{sign}(H_{ji})q_i$. If $p(j|i) > 1$ then the corresponding integer number of walkers are realized deterministically, and the fractional part stochastically \cite{booth_fermion_2009}.
		
		\item \textit{Clone/Death:} For each walker on a given site, a death event is sampled with probability
		
		\begin{equation}
		p_{death}(i) = \Delta\tau(H_{ii}-S)\,.
		\end{equation}
		
		If $p_{death} > 0$, the walker is removed from the population. If $p_{death} < 0$, a walker of opposite sign is created. In case $|p_{death}| > 1$, the integer part of $p_{death}$ is realized deterministically, and the fractional part stochastically.
		
		\item \textit{Annihilation:} pairs of walkers of opposite sign residing on the same basis state are annihilated. Therefore, at the end of each time step, each state is solely occupied by walkers having the same sign. 
		
	\end{enumerate}
	
	One of the most significant advantages of FCIQMC is due to the aforementioned annihilation procedure. This step does not alter the evolution of the ground state, but was shown to be crucial in systems with sign problem \cite{spencer_sign_2012}. The sign problem in FCIQMC simulation is manifested as the fast growth of an unphysical solution dominating the ground-state result. The annihilation procedure can suppress this growth and allow the simulation to converge to the physical solution, but only if a minimal and system-dependent walker population is present. Henceforth, building on the massively parallel nature of the method, a computationally efficient implementation can offer an insight into the study of systems with severe sign problem.

	\section{\label{sec:ddqmc} Driven-dissipative Quantum Monte Carlo}
	\main We describe now how the dynamics of open quantum systems following the Liouville-von-Neumann equation can be cast into a Monte Carlo algorithm. 
	
	\subsection{Theory}
	
	\main The general problem we aim to solve is that of quantum system with several degrees of freedom, in the presence of external driving fields and Markovian coupling to the environment. The evolution of the steady matrix $\hat{\rho}$ is then governed by the Liouville-von-Neumann master equation \cite{gardiner_quantum_2004}
	
	\begin{equation}\label{lindblad}
	\frac{d\hat{\rho}}{d t} = \mathcal{L}(\hat{\rho}) = -i[\hat{H}, \hat{\rho}] + \sum\limits_{i}^{}\mathcal{L}_i(\hat{\rho})\,.
	\end{equation}
	
	\noindent The dissipative part of the dynamics is described by
	
	\begin{equation}
	\sum\limits_{i}^{}\mathcal{L}_i(\hat{\rho})=- \sum\limits_{i}^{}\frac{\gamma_i}{2}\left[\left\{\hat{F}_i^{\dagger}\hat{F}_i, 
	\hat{\rho}\right\}-2\hat{F}_i\hat{\rho}\hat{F}_i^{\dagger}\right]\,,
	\end{equation}
	
	\noindent where $\hat{F}_i$ are the jump operators, characterizing the transitions induced by the environment, and $\gamma_i$ are the corresponding transition rates. Contrary to the Hamiltonian, the Liouville superoperator is not hermitian. Dissipative systems evolve under a one parameter semi-group $(e^{\mathcal{L}t}, t>0)$, generated by the Liouvillian, resulting in a time evolution which is no longer unitary. Due to its non-hermiticity, $\mathcal{L}$ has complex eigenvalues with negative real part. It can be shown that the density matrix, under very general assumptions will evolve into an asymptotic steady state, corresponding to the zero eigenvalue of $\mathcal{L}$ \cite{albert_symmetries_2014-1}. \\
	\main By introducing an additional shift into eq.~\eqref{lindblad} for a diagonal population control,
	\begin{equation}\label{lindShift}
	\frac{d\hat{\rho}}{d t} = \mathcal{L}(\hat{\rho}) -S\hat{\rho}=\widetilde{\mathcal{L}}(\hat{\rho})\,.
	\end{equation}
	\noindent Eq.~\eqref{lindShift} can be stochastically sampled similarly to the Hamiltonian case and the NESS is obtained as a Monte Carlo average of the long time limit. As in the case of FCIQMC, we take the first-order Euler approximation,
	\begin{equation}\label{Euler}
	\hat{\rho}(t+\Delta t) = \hat{\rho}(t) + \widetilde{\mathcal{L}}(\hat{\rho})\cdot\Delta t\,,
	\end{equation}
	and we introduce a set of walkers which now sample the amplitudes of basis operators $|\phi_i\rangle\langle\phi_j|$, from now on referred as ``configurations". \\
	\main The stochastic sampling of the unnormalized density matrix gives access to the expectation value of any quantum mechanical observable. The expectation value of observable $\hat{O}$ at a given instant in time is computed by
	
	\begin{equation}
	\langle\hat{O}\rangle(t) = \frac{\mathrm{Tr}[\hat{O}\hat{\rho}(t)]}{\mathrm{Tr}[\hat{\rho}(t)]}=\frac{\sum_{i,j}^{}\rho_{ij}(t)O_{ji}}{\sum_{i}^{}\rho_{ii}(t)}\,.
	\end{equation}
	
	\noindent Once the simulation has asymptotically approached the steady state -- i.e. the shift $S$ reached the steady state eigenvalue of the Liouvillian, $S=0$ -- the numerator and the denominator can be averaged separately over a sufficiently large number of iteration steps. \\
	
	\subsection{Multinomial formalism}
	\label{multiform}
	\main The original FCIQMC sampling protocol was described, e.g. in Ref. \cite{booth_fermion_2009}. Here we developed a variant which optimizes the computational cost of the evolution generated by off-diagonal Hamiltonian elements. In the followings we will refer to this variant as multinomial formalism. \\
	\main In the original scheme, in order to perform the stochastical evolution induced by the off-diagonal $H_{ij}$ elements in eq.~\eqref{dyn_eq}, the algorithm requires a loop over the entire walker population at each time step. This method becomes computationally heavy as the walker population increases.\\
	\main Here we introduce an alternative strategy for the spawning generation. Let $p(j|i)$ be the probability of choosing the $j$-th child starting from site $i$
	
	\begin{equation}
	p(j|i) = \frac{|H_{ji}|\Delta\tau}{\sum_k |H_{ki}|\Delta\tau} = \frac{|H_{ji}|\Delta\tau}{P_{tot}}\,.
	\end{equation}
	
	\noindent Then the number of actual spawning events $N_{sp}^i$ occurring for $N_i$ walkers residing on site $i$ is determined by a stochastic process following a binomial distribution
	
	\begin{equation}
	f(N_{sp}^i;N_i,P_{tot}) = \frac{N_i!}{N_{sp}^i!(N_i-N_{sp}^i)!}P_{tot}^{N_{sp}^i}(1-p)^{N_i-N_{sp}^i}\,.
	\end{equation}
	
	\noindent Then the $N_{sp}^i$ walkers are divided into groups $\{M_1 \ldots M_l\}$, where $l$ is the number of states connected to the starting one by a non-zero Hamiltonian element. For each group, $M_j$ children are spawned to the $j$-th site with sign $q_j = \mathrm{sign}(H_{ji})q_i$. The set of integers $\{M_j\}$ is drawn randomly following the multinomial distribution
	
	\begin{equation}
	\begin{array}{lcl}
	f_M(M_1 \ldots M_l;N_{sp}^i,p(1|i) \ldots p(l|i)) =\\
	
	\vspace*{0.2cm}\hspace*{1cm}
	=\frac{N_{sp}^i!}{M_1!\cdots M_l!}\hspace*{0.1cm}p(1|i)^{M_1}\times\cdots\times p(l|i)^{M_l}\,.
	
	\end{array}
	\end{equation}
	
	Therefore, in each time step we perform a loop over the currently populated basis states rather than the whole walker population. In systems with local coupling the Hamiltonian is represented by a highly sparse matrix, and a computationally effective state representation makes the extra memory allocation negligible (for an occupied site $i$ it is necessary to store all the possible connected states with the corresponding probabilities). Efficient algorithms for binomial and multinomial random number generation are also present in the literature \cite{kachitvichyanukul_binomial_1988,davis_computer_1993,hormann_generation_1993}.

	\subsection{Algorithm}
	
	\main The dynamics of the walker population is determined by a set of rules designed to stochastically sample eq.~\eqref{Euler}. However, the elements of the density matrix are complex valued. Similarly to \cite{booth_towards_2013}, we can sample a complex density matrix with two types of walkers, respectively for the real and imaginary parts. If the density matrix is expressed in vectorized form, the shifted Liouvillian superoperator can be expressed im matrix form using Kronecker products as \cite{jakob_variational_2003}
	
	\begin{equation}
	\begin{split}
	\mathcal{\widetilde{L}} = &-i(\mathbb{1}\otimes\hat{H}-\hat{H}^T\otimes\mathbb{1}) - S\cdot\mathbb{1}\otimes\mathbb{1}\\
	&+\sum_{i}^{}\frac{\gamma_i}{2}(2\hat{F}_i^*\otimes\hat{F}_i-\mathbb{1}\otimes\hat{F}_i^\dagger\hat{F}_i-\hat{F}_i^T\hat{F}_i^*\otimes\mathbb{1})\,.
	\end{split}
	\end{equation}
	
	\noindent Then eq.~\eqref{lindShift} can be written in the form of
	
	\begin{equation}\label{eq:elmeq}
	\frac{d\rho_{ij}}{d t} = \widetilde{\mathcal{L}}_{ij}^{ij}\rho_{ij} + 
	\sum_{l,m \neq i,j}^{}\widetilde{\mathcal{L}}_{ij}^{lm}\rho_ {lm}\,,
	\end{equation}
	
	\noindent where $\widetilde{\mathcal{L}}_{ij}^{lm}$ are the matrix elements of the superoperator. Here $\rho_{ij}$ represents the now complex valued population on a given configuration $|\phi_i\rangle\langle\phi_j|$.\\
	\main For the sampling protocol we use the multinomial formalism, introduced in Sec.~\ref{multiform}. Let us refer to this scheme as a function
	\begin{equation}
	M_{lm}=\mathbf{Multinomial}(A_{ij}^{lm})\,,
	\end{equation}
	which returns randomly drawn number of walkers spawned from configuration $ij$ to $lm$ given the matrix element connecting them.
	At each time step, we loop over the list of currently occupied configurations and perform the following steps:
	
	\begin{enumerate}[label=(\roman*)]
		
		\item \textit{Spawning:} Consider the complex walker population residing on $\rho_{ij}$ and perform spawning to all the connected configurations. The real ($\Re$) and imaginary ($\Im$) parts of $\widetilde{\mathcal{L}}_{ij}^{lm}$ are considered in turn and two spawning attempts are realized respectively for real and imaginary parents. The number of walkers spawned to each $\rho_{lm}$ are determined by the multinomial formalism.\\\\
		For real parents
		
		\begin{equation}
		\left[
		\begin{array}{l}
		N_{lm}^\Re = \mathbf{Multinomial}(\Re(\widetilde{\mathcal{L}}_{ij}^{lm})) \\
		
		\mathrm{sign}=\mathrm{sign}(\Re(\rho_{ij})\Re(\widetilde{\mathcal{L}}_{ij}^{lm}))
		\end{array}
		\right.
		\end{equation}
		
		\begin{equation}
		\left[
		\begin{array}{l}
		N_{lm}^\Im = \mathbf{Multinomial}(\Im(\widetilde{\mathcal{L}}_{ij}^{lm})) \\
		
		\mathrm{sign}=\mathrm{sign}(\Re(\rho_{ij})\Im(\widetilde{\mathcal{L}}_{ij}^{lm}))
		\end{array}
		\right.
		\end{equation}
		
		and for imaginary parent walkers
		
		\begin{equation}
		\left[
		\begin{array}{l}
		N_{lm}^\Re = \mathbf{Multinomial}(\Im(\widetilde{\mathcal{L}}_{ij}^{lm})) \\
		
		\mathrm{sign}=-\mathrm{sign}(\Im(\rho_{ij})\Im(\widetilde{\mathcal{L}}_{ij}^{lm}))
		\end{array}
		\right.
		\end{equation}
		
		\begin{equation}
		\left[
		\begin{array}{l}
		N_{lm}^\Im = \mathbf{Multinomial}(\Re(\widetilde{\mathcal{L}}_{ij}^{lm})) \\
		
		\mathrm{sign}=\mathrm{sign}(\Im(\rho_{ij})\Re(\widetilde{\mathcal{L}}_{ij}^{lm}))
		\end{array}
		\right.
		\end{equation}
		
		where $N_{lm}^\Re$ and $N_{lm}^\Im$ are the number of real and imaginary walkers being spawned to configuration $\rho_{lm}$ and '$\mathrm{sign}$' indicates the sign of the progeny.
		
		\item \textit{Clone/Death:} This step is required as a real (imaginary) walker can produce an imaginary (real) walker on the same configuration. The spawning occurs on-site with a population determined by the binomial distribution. The real and imaginary parts of $\widetilde{\mathcal{L}}_{ij}^{ij}$ are considered in turn and two spawning attempts are realized respectively for the real and imaginary population.
		
		\item \textit{Annihilation:} on a given site the real and imaginary population are considered in turn, and pairs of walkers having opposite signs are removed from the simulation.	
		
	\end{enumerate}
	
	\noindent Unlike in FCIQMC, here the target value of the diagonal shift $S$ is known. In the infinite time limit the master equation is assumed to drive the density matrix into the steady state, which by definition belongs to the zero eigenvalue of the Lindbladian. Therefore, the time evolution of the shift $S$ will indicate if convergence is reached and one can start gathering statistics on the observables. As before, the shift is slowly adjusted in order to maintain a constant walker population. Since estimators for most operators of interest only receive contributions from walkers on or near the diagonal elements, we chose to control the amount of population distributed along the diagonal of the density matrix. The value of the shift is then adjusted according to the familiar shift-update algorithm implemented in FCIQMC \cite{booth_fermion_2009} calculations
	
	\begin{equation}
	S(t) = S(t-\Delta t)+ \frac{\delta}{\Delta t}\mathrm{log}\left(\frac{N_w(t)}{N_w(t-\Delta t)}\right)\,,
	\end{equation}
	
	\noindent where $\delta$ is a damping parameter, and $N_w$ is the total weight of real walkers residing on diagonal density matrix elements. The method does not have a built-in constraint on the diagonal elements being real. The value of the imaginary part fluctuates around zero and its expectation value naturally vanishes during the simulation. 
	
	\section{\label{sec:initiator} Initiator approach \& importance sampling}
	\main The algorithm described in Section~\ref{sec:ddqmc} allows the stochastical sampling of the steady state density matrix of open quantum systems whose dynamics follows the Liouville-von-Neumann equation. However, with the increase in the number of configurations, the walker population tends to become dilute, resulting in a situation where the simulation contains only a few walkers per basis operator. This leads to an increased statistical error thus reducing dramatically the accuracy of the sampling. In order to address this issue one needs to increase the walker number in the system which -- with large system sizes -- becomes computationally unfeasible. \\
	\main In order to overcome this issue two different methods were introduced: initiator approach and importance sampling. Each of the techniques reduces the minimal required walker population by decreasing the number of simultaneously occupied configurations, however the strategy of selecting the configurations to be sampled is fundamentally different. The initiator approach allows the significant configurations to emerge naturally during the simulation, whilst importance sampling gives the possibility to drive the walker population to a selected subset of presumably relevant configurations. \\
	\main These methods can improve the sampling quality with great success, however, they have to be applied carefully, as both introduce a bias on the result.
	
	\subsection{Initiator approach}
	\begin{figure}[t!]
		\centering\includegraphics[width = 0.5\textwidth]{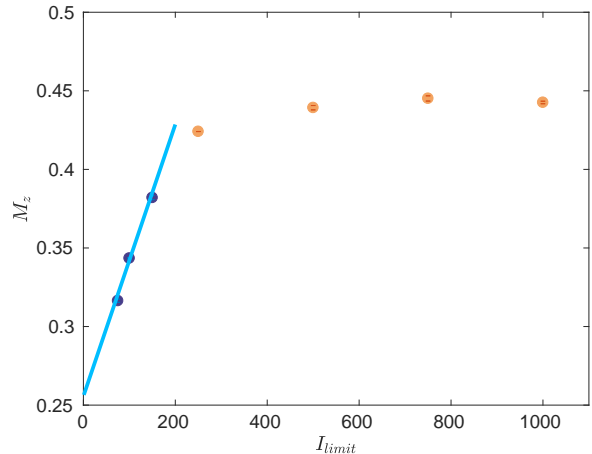}
		\vspace*{-1.3cm}
		\caption{\label{fig:init} The initiator approach used to extrapolate the $M_y$ magnetization in case of the $4\times4$ dissipative XYZ Heisenberg-model. Parameters of the model: $J_x/\gamma = 0.225$, $J_y/\gamma=0.335$, $J_z/\gamma = 0.25$, $h=0.1$, $\theta=0$. Parameters of the simulation: $p=2.5$, with a population of $10^6$ walkers. The straight line is a linear extrapolation of the lowest initiator values.}
	\end{figure}

	\main Our initiator approach is based on an FCIQMC adaptation detailed in \cite{cleland_communications:_2010}. For the newly spawned walkers an additional survival criteria is introduced which can dramatically reduce the population required to reach convergence. Some of the basis operators are tagged as 'initiator' which have the ability to spawn progeny onto unoccupied basis states. However, progeny of the non-initiator states can only survive if they spawn to basis operators which were previously occupied or to diagonal elements. All the diagonal basis operators are initiators by definition, and during the simulation a basis element might become initiator, if its population exceeds a preset value ($I_{limit}$).\\
	\main This method results in a series of systematically improvable approximations which will tend to the original algorithm in three limits:
	
	\begin{enumerate}[label=(\roman*)]
		
		\item with decreasing $I_{limit}$ every basis element will become initiator. All the progeny survives regardless of the parent state, which is equivalent to the original method; 
		
		\item with increasing total population all basis element will acquire walkers, therefore, all spawned children will survive regardless of the flag of the parent state;
		
		\item extending the initiator space by definition will result again in all the basis operators becoming initiators, consequently all progeny will survive.
		
	\end{enumerate}
	
	\main Setting an initiator limit, introduces a dynamical truncation on the available basis operators, leading to a biased result. In order to compute the unbiased expectation values, we progressively decrease the initiator limit in different simulations and fit the estimated expectation values, thus extrapolating the value in the limit $I_{limit}\rightarrow 0$ (Figure~\ref{fig:init}).

	\subsection{Importance sampling}
	\begin{figure}[t!]
		\includegraphics[width = 0.5\textwidth]{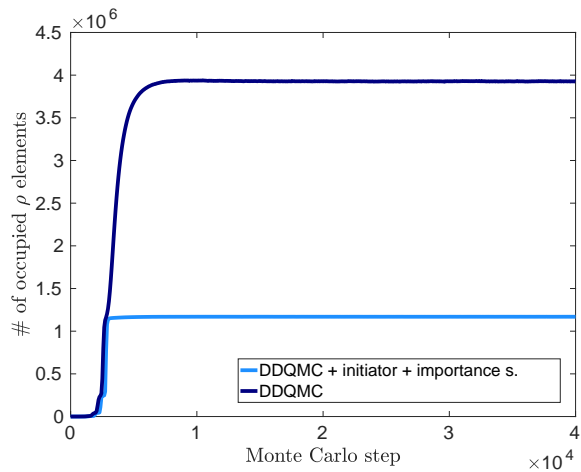}
		\caption{\label{fig:impscale} The amount of simultaneously occupied density matrix elements, with and without using the initiator approach and importance sampling in the case of the $4\times4$ XYZ Heisenberg lattice. Parameters of the model: $J_x/\gamma = 0.225$, $J_y/\gamma=0.225$, $J_z/\gamma = 0.25$, $h=0.1$, $\theta=0$. Parameters of the simulation: $p=1.5$, $I_{limit}=25;75$ with a population of $10^8$ walkers.}
	\end{figure}
	
	\main We start by identifying the basis elements whose sampling needs to be improved. Then a straightforward way to do so is by reducing the probability of spawning out these configurations. Walkers that do reside on unessential elements are given a correspondingly larger weight, hence the expectation values of the observables will be unchanged. We define the following simple importance sampling procedure. \\
	\main The evolution of the density matrix in the DDQMC formalism follows eq.~\eqref{eq:elmeq}. In order to associate weights to the matrix elements depending on their importance we multiply them a factor $w_{ij}$
	
	\begin{equation}
	\rho_{ij} \rightarrow \widetilde{\rho}_{ij}=w_{ij}\rho_{ij}\,.
	\end{equation}
	
	\noindent The time evolution of the importance-sampled density matrix then becomes
	
	\begin{equation}
	\frac{d\widetilde{\rho}_{ij}}{d t}=\widetilde{\mathcal{L}}_{ij}^{ij}\widetilde{\rho}_{ij} + 
	\sum_{l,m \neq i,j}^{}\left(w_{ij}\widetilde{\mathcal{L}}_{ij}^{lm}\frac{1}{w_{lm}}\right)\widetilde{\rho}_ {lm}\,,
	\end{equation}
	\noindent which is fully analogous to eq.~\eqref{eq:elmeq} and can be simulated by the DDQMC method. The expectation value of an observable $\hat{O}$ is thus
	\vspace*{-0.2cm}
	\begin{equation}
	\langle\hat{O}\rangle=\frac{\sum_{ij}^{}\frac{\widetilde{\rho}_{ij}}{w_{ij}}O_{ji}}{\sum_{i}^{}\widetilde{\rho}_{ii}}\,.
	\end{equation}

	\main In this work we introduce a single importance sampling parameter $p>0$, and give all the off-diagonal elements a weight $w_{ij}=e^{-p}$. Meanwhile the diagonal coefficients are not altered. This strategy focuses on sampling the diagonal density matrix elements and gives an easy access to tune the strength of the importance sampling. \\
	\main Fig.~\ref{fig:impscale} shows the amount of simultaneously occupied density matrix elements before and after using the initiator approach and importance sampling given the XYZ Heisenberg lattice model that we describe later.

	\section{\label{sec:results} Results}
\main In order to demonstrate the effectiveness of DDQMC, we simulated the two-dimensional spin-1/2 XYZ Heisenberg lattice in the presence of a dissipating channel which tends to relax each spin into the $|s_z=-1/2\rangle$ state. 

\begin{figure}[t!]
	
	\centering\includegraphics[width = 0.5\textwidth]{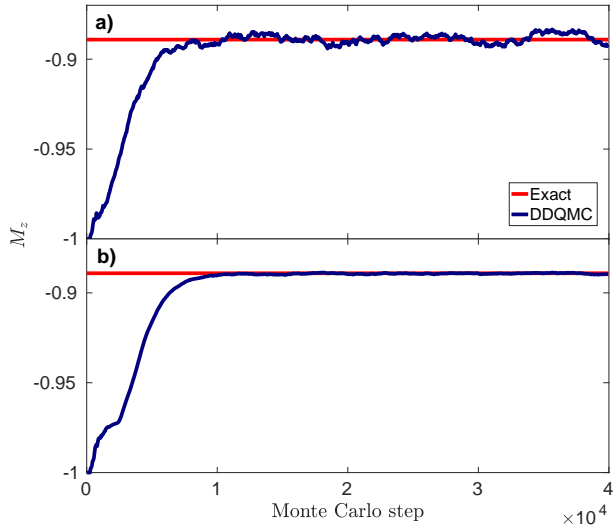}
	\caption{\label{fig:stocherr} The exact and the DDQMC magnetization values for the $3\times3$ dissipative XYZ Heisenberg lattice with periodic boundary condition. The coupling parameters are $J_x/\gamma = 0.225$, $J_y/\gamma = 0.335$ and $J_z/\gamma = 0.25$. The diagonal population was limited to (a) $50^4$ and (b) $20^6$ walker.}
\end{figure}

The model follows the Liouville-von-Neumann equation and the Hamiltonian is governed by ($\hbar=1$)

\begin{eqnarray}
\hat{H} = \sum\limits_{\langle i,j\rangle}^{}\left(J_x\hat{S}_i^x\hat{S}_j^x + J_y\hat{S}_i^y\hat{S}_j^y + J_z\hat{S}_i^z\hat{S}_j^z\right) \\
\frac{\mathrm{d}\hat{\rho}}{\mathrm{d}t} = -i[\hat{H}, \hat{\rho}] - \frac{\gamma}{2}\sum\limits_j^{}\left[\left\{\hat{S}_j^+\hat{S}_j^-, 
\hat{\rho}\right\}-2\hat{S}_j^-\hat{\rho}\hat{S}_j^+\right]
\end{eqnarray}

\noindent where $\hat{S}_i^\alpha$ are the spin operators matrices acting on the $i$-th spin, $J_\alpha$ are the coupling constants between nearest neigbour spins, $\gamma$ is the dissipation rate, and $\hat{S}_j^{\pm} = \hat{S}_j^x\pm i\hat{S}_j^y$. \\
\main Recently, the system has attracted significant interest since the competition between the coherent Hamiltonian dynamics and the incoherent spin flips leads to a dissipative phase transition. Due to the anisotropy, the Hamiltonian part induces a nonzero spin expectation value on the $xy$ plane, while dissipation would drive each site to the spin down state. This competition leads to a phase transition from a paramagnetic phase (with no magnetization in the \textit{xy} plane) to a magnetically organized phase (which presents a finite polarization in the \textit{xy} plane). Both the Gutzwiller mean-field theory \cite{lee_unconventional_2013,jin_cluster_2016,casteels_gutzwiller_2017} and the corner space renormalization method \cite{rota_critical_2017,rota_dynamical_2017} predicts this dissipative phase transition.\\
\main In order to study the model we chose three different lattice sizes: $2\times2$, $3\times3$ and $4\times4$. The first two sizes are small enough to derive an exact numerical solution of the master equation in the steady state, thus allowing a direct check of the accuracy of our DDQMC results. In case of the $4\times4$ lattice, the magnetization is compared to those obtained by Monte Carlo wave function technique.

\subsection{Magnetization in the steady-state}

\main The steady-state magnetization per site is defined as

\begin{equation}
M_z = \frac{1}{N}\sum_{i=1}^{N}\mathrm{Tr}(\hat{\rho}\hat{\sigma}_i^z),
\end{equation}

\noindent where $N$ is the number of lattice sites. \\
\begin{figure}[t!]
	\centering\includegraphics[width = 0.5\textwidth]{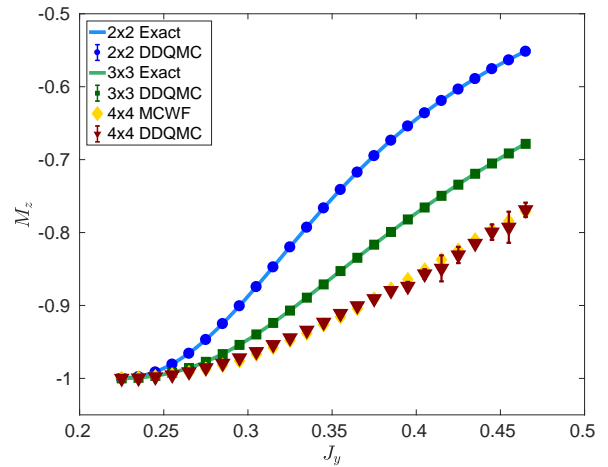}
	\caption{\label{fig:magnet} The magnetization $M_z$ per site as a function of the normalized coupling parameter $J_y/\gamma$ for different lattice sizes. The other coupling parameters are $J_x/\gamma = 0.225$ and $J_z/\gamma = 0.25$. The exact ($2\times2$ and $3\times3$) and numerical ($4\times4$) results are plotted for comparison.}
\end{figure}
\main Figure \ref{fig:stocherr}\textcolor{blue}{(a)} shows the magnetization of the $3\times3$ lattice as a function of the Monte Carlo iteration step with a diagonal population of $50^4$ walkers. The exact solution obtained by directly solving the linear system is also plotted. Increasing the diagonal population to $20^6$ reduces the statistical error as seen in the corresponding result in Fig.~\ref{fig:stocherr}\textcolor{blue}{(b)}. \\
\main In Fig. \ref{fig:magnet}, we present the magnetization per site $M_z$ as a function of the normalized coupling constant $J_y/\gamma$ for square lattices of different size. The exact and MCWF solutions are also plotted and are in agreement with the results obtained by the DDQMC algorithm.

\subsection{Angularly-averaged susceptibility}

\main Following the scheme presented in \cite{rota_critical_2017}, we study the system in the presence of an applied magnetic field in the \textit{xy} plane

\begin{equation}
\hat{H}_{ext} = \sum_{i}^{}h(\cos(\theta)\hat{\sigma}_i^x + \sin(\theta)\hat{\sigma}_i^y)\,.
\end{equation}

The linear response is then summarized in the $2\times2$ susceptibility tensor
\begin{equation}
\chi_{\alpha\beta} = \frac{\partial M_\alpha}{\partial h_\beta}\Bigr|_{\substack{h=0}} \,,
\end{equation}
\noindent with $\alpha,\beta=x,y$. 

It is convenient to calculate one single quantity, the angularly-averaged susceptibility

\begin{equation}
\chi_{av}=\frac{1}{2\pi}\int_{0}^{2\pi}\mathrm{d}\theta\frac{\partial |\vec{M}(h,\theta)|}{\partial h}\Bigr|_{\substack{h=0}}\,,
\end{equation}

\noindent where

\begin{equation}
\frac{\partial |\vec{M}(h,\theta)|}{\partial h}\Bigr|_{\substack{h=0}} = 
\left|\left(
\begin{matrix}
\chi_{xx}\cos(\theta)+\chi_{xy}\sin(\theta) \\
\chi_{yx}\cos(\theta)+\chi_{yy}\sin(\theta) 
\end{matrix}
\right)\right|\,.
\end{equation}

\begin{figure}[t!]
	\centering\includegraphics[width = 0.5\textwidth]{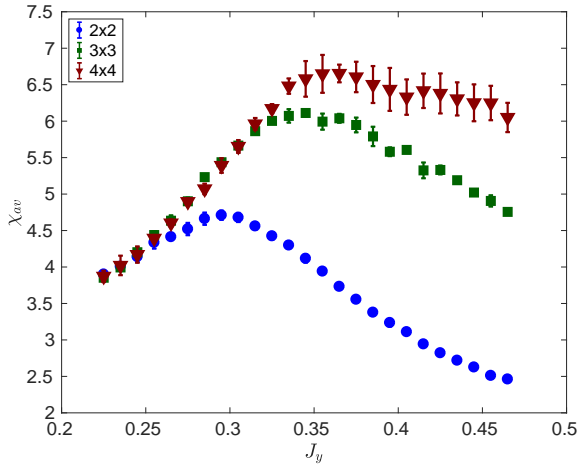}
	\caption{\label{fig:susc} The angle-averaged susceptibility $\chi_{av}$ per site as a function of the normalized coupling parameter $J_y/\gamma$ for different lattice sizes. The other coupling parameters are $J_x/\gamma = 0.225$ and $J_z/\gamma = 0.25$. Each point on the plot was determined by 21 simulations, which corresponds to 525 calculations per lattice size. For each point, we considered $3+3+1$ values of the applied field ($3$ for each in-plane direction and $1$ with no external field), and for each setting an extrapolation over three different initiator values was carried out. }
\end{figure}

For a more complete derivation readers are referred to \cite{rota_critical_2017}. In Fig. \ref{fig:susc}, we present the angularly-averaged susceptibility $\chi_{av}$ as a function of the normalized coupling parameter $J_y/\gamma$ for different lattice sizes.

\main The magnetic susceptibility for the different lattice sizes exhibits a peak of increasing height which qualitatively corresponds to the results obtained in \cite{rota_critical_2017}.\\

	\section{\label{sec:out} Outlook}
	\main The DDQMC method presented here, constitutes a basic PMC approach to the non-equilibrium steady state of open quantum systems. As such, it contains only the basic building blocks of the PMC method, and its effectiveness may be considerably improved by introducing any of the several tools that are common in other PMC schemes. Here, we describe as an outlook three such possible improvements. \\
	\main The first possible improvement consists in the implementation of a mixed-estimator scheme, in analogy with the one used in projector and diffusion Monte Carlo to find the ground state of Hamiltonian systems \cite{kolorenc_applications_2011,foulkes_quantum_2001,umrigar_observations_2015}. Here, a possible mixed estimator strategy may consist in formally carrying out an exact real-time evolution, starting from a DDQMC sampled density matrix. More specifically, let us assume that at time $t$ the current DDQMC sample of the density matrix is $\hat\rho(t)$. We can formally apply the exact time evolution for an additional time $T$ and then evaluate the expectation value of an observable $\hat O$ at time $t+T$ as
	\begin{equation}
	\langle\hat O\rangle=\frac{\mbox{Tr}(\hat O e^{{\cal L}T}\hat\rho(t))}{\mbox{Tr}(e^{{\cal L}T}\hat\rho(t))}=\frac{\mbox{Tr}(\hat O e^{{\cal L}T}\hat\rho(t))}{\mbox{Tr}(\hat\rho(t))}\,,
	\label{Mixed}
	\end{equation}
	where the second equality results from the trace preserving character of the time evolution. In the limit $T\rightarrow\infty$, Eq. (\ref{Mixed}) provides the steady-state expectation value independently of the actual value of the density matrix $\hat\rho(t)$, when assuming that a unique steady state exists. A mixed estimator strategy would then consist in building a ``trial'' observable $\hat O_T$ which can still be efficiently computed element-wise, and such that $\hat O_T\simeq\hat O_H(T)= \hat O e^{{\cal L}T}$. Here, $\hat O_H(t)$ represents the Heisenberg picture of the observable $\hat O$ and, for time-independent Liouvillian maps, it obeys the adjoint quantum master equation $\frac{d\hat O_H(t)}{dt}={\cal L}^\dagger \hat O_H(t)$ \cite{breuer_theory_2002}. Hence, the mixed estimator approach in the present case would require the knowledge of an approximate time dependence for $\hat O_H(t)$, which may be obtained, for instance, from a time-dependent variational principle \cite{haegeman_time-dependent_2011,mascarenhas_diffusive-gutzwiller_2017} applied to a separable or short-range-correlated ansatz for the observable. \\
	\main A second improvement would consist in using a ``guiding density matrix'' for the importance sampling. A natural choice for such a guiding density matrix would again be a variational ansatz, as the variational principle for the NESS is now well established [6], and some variational approaches have already been developed \cite{weimer_variational_2015,mascarenhas_matrix-product-operator_2015,cui_variational_2015}.\\
	\main Finally, the present scheme is based on the Euler method (\ref{Euler}) for the numerical solution of the time-dependent master equation. The Euler method is a first order method in the time step, and is only stable if $\Delta t$ is chosen to be smaller than the inverse of the full spectral width of the master equation. In PMC, several approaches have been proposed to sample a higher-order discrete-timestep propagator \cite{umrigar_diffusion_1993}, or even the exact one \cite{schmidt_greens_2005}. While a similar approach would be highly beneficial to FCIQMC and DDQMC, the question is still open, whether higher-order propagators may be efficiently sampled within the spawn-annihilation sampling protocol characterizing these Monte Carlo methods.\\

	\section{\label{sec:disc} Conclusions}
We have introduced a quantum Monte Carlo approach to open many-body quantum systems, called DDQMC. The method is based on the FCIQMC algorithm exploiting the analogy between the long-time dynamics of the Lindbladian master equation and the imaginary-time Schrödinger equation. DDQMC allows direct sampling of the steady state density matrices in any discrete basis set, and in all cases studied it has proven to be accurate. \\
\main  DDQMC, as FCIQMC, uses an annihilation procedure which helps to alleviate the sign problem. The introduction of the initiator approach and importance sampling can lead to a significant improvement in the statistical accuracy and reduce the required walker population. The validity of the method was proven by investigating a dissipative phase transition on the two-dimensional Heisenberg-model. The defining feature of DDQMC is that it samples the whole density matrix and it does not introduce a truncation in Hilbert-space. Experience showed that the applicability of the code does not solely depend on the system size, but also on the correlations characterizing the steady state. The application presented in this work is a proof of principle, demonstrating the possibility to stochastically sample the Liouville-von-Neumann equation in a finite difference approximation. DDQMC holds promise as a powerful tool in the study of open quantum systems.

\begin{acknowledgments}
	We acknowledge enlightening discussions with Markus Holzmann and Eduardo Mascarenhas. We are indebted to Hugo Flayac for having provided the MCWF simulations used to benchmark the present results. This work was supported by the Swiss National Science Foundation through Project No. 200021\_162357.
\end{acknowledgments}

	
	
	\nocite{*}
	\bibliography{ddqmcBib}
	
\end{document}